\documentclass[12pt]{article}
 
 \usepackage{amssymb,amsmath,enumerate,amsthm}

 \theoremstyle{definition} 
 \newtheorem{remark}{Remark}[section]
 \newtheorem{example}{Example}[section]
 \newtheorem{definition}{Definition}[section]
 
\title{Background-Independence}
\author{Gordon Belot\footnote{Forthcoming in \emph{General Relativity and Gravitation.}}  \\ University of Michigan \\ belot@umich.edu }

\begin{document}

\maketitle 
\begin{abstract}  \noindent Intuitively speaking, a classical field theory is back\-ground-independent if the structure required to make sense of its equations is itself subject to dynamical evolution, rather than being imposed \emph{ab initio}. The aim of this paper is to provide an explication of this intuitive notion. Background-independence is not a not formal property of theories: the question whether a theory is background-independent depends upon how the theory is interpreted. Under the approach proposed here, a theory is fully background-independent relative to an interpretation if each physical possibility corresponds to a distinct spacetime geometry; and it falls short of full background-independence to the extent that this condition fails. 
\end{abstract}

\newpage

\section{Introduction}

In in pre-relativistic physics, as in common sense, space and time provide a fixed scaffolding or stage against which physical systems evolve. 

\begin{quote} Ask an intelligent man who is not a scholar what space and time are, and he will perhaps answer as follows. If we imagine all physical things, all stars, all light taken out of the universe, what then remains is something like a giant vessel without walls called `space.' With respect to what is happening in the world, it plays the same role as the stage in a theater performance. In this space, in this vessel without walls, there is an eternally uniformly occurring tick-tock \ldots that is `time.' Most natural scientists, up to the present, had this conception about the essence of space and time \ldots. (Einstein \cite[item 44a]{:2002yq}) \end{quote}
It is often suggested that what makes general relativity (and theories of its ilk) special is their \emph{background-independence}---their treatment of space and time as among the actors in the theory rather than as a fixed stage upon which the action unfolds. 

`Background-independence' has no precise and fixed meaning. But according to the sense in which the term is used here, general relativity (in its usual formulation and under its usual interpretation) is a paragon of background-independence, while the theory of a Klein--Gordon field in Minkowksi spacetime is a paragon of background-dependence (whether or not the spacetime metric is treated as a dynamical variable).\footnote{In fact, of course, the term `background-independence' is used in many ways. The use in the present discussion is very close to that found in, e.g., \cite{Giulini:qm} and \cite{Rovelli:2004uh}. %[{esp. pp. 10 and 31}]
Sometimes, however, `background-independence' is used as a synonym for `generally covariant' \cite{Sorkin:2002ab}%affixed[p. 698]{see, e.g.,}
---under this usage, a theory of a Klein--Gordon field propagating against a Minkowski metric counts as background-independent if the spacetime metric is treated as a variable in the equations of motion. Again, `background-independence' is sometimes used as a name for a sweeping form of relationalism \cite{Smolin:2006yq}%affixed[\S\S 3 and 4]{see, e.g.,}
---and then even general relativity is not fully background-independent, because the topological structure of spacetime and the signature of the metric are fixed \emph{ab initio}. For another approach, see \cite{Gryb:2010rz}.} 

Intuitively, a background-independent theory must be generally covariant---any structure that breaks general covariance is background structure. And if a theory is generally covariant without being background-independent, then its general covariance must in some sense be artificial or dispensable---one expects general covariance to be broken in any formulation in which the background is made explicit. So, speaking very roughly and intuitively, a theory is background-independent if and only if its most perspicuous formulation is generally covariant (but see the discussion of the  Einstein--Maxwell theory in Section \ref{ssecMatter} below).  

The question whether future theories of physics can be expected to be background-independent and the question whether various proposed approaches to quantum gravity live up to this expectation have been very widely discussed in recent years. The goal of the present discussion is to offer an explication of the intuitive notion of background-independence as it applies to classical field theories with the hope that clear standards of background-independence will provide a framework to structure such debates.  

It might be thought that the notion of an absolute object \cite{Anderson:1964nr,Anderson:1967tu,Earman:1974te,Friedman:1973nw,Friedman:1983rc} should play a crucial role here. For an absolute object of a theory is a field that is in a certain sense \emph{the same} in each solution---and so it might naturally be suggested that a theory is background-independent if and only if it features no absolute objects. However, this suggestion runs into serious difficulties  (see Remark \ref{remAbs} below). A different one is offered in its place---roughly speaking, that a theory is fully background-independent if alteration of the physical degrees of freedom always implies an alteration in geometry; and that a theory falls short of full background-independence to the extent that this condition fails. 

The analysis is intended to apply only to theories of the universe as a whole in which all interactions between systems are taken into account. The clearest cases of background-dependent theories are theories featuring fixed (i.e., solution-independent) fields. Such fields can play a number of roles in a theory---but in practice, they are typically used to represent geometry, fixed sources, or external fields. Plausibly, in a theory in which background structure is present without being encoded in fixed fields, the background still plays one of these same roles. By restricting attention to theories of the universe as a whole in which all interactions are taken into account, the possible roles of background are narrowed down to a single one---geometry. 

In Section \ref{secFields} below, a framework for talking about field theories and their symmetries is sketched. Section \ref{secEx} includes a preliminary discussion of background-independence illustrated by several examples. 
Sections \ref{secGeom} and \ref{secCount} develop the notions of geometrical degrees of freedom and of physical degrees of freedom that are required for the present approach. Various gradations of background-(in)dependence are defined in Section \ref{secBI}. Features and limits of these notions are discussed in Section \ref{secDisc}. 

\emph{Conventions.} All Lie groups are taken to be connected. Where $g$ denotes a metric, $\mathbf{Ein}[g],$ $\mathbf{Weyl}[g],$ $\mathbf{Riem}[g],$ $\mathbf{Ricci}[g],$ and $\mathbf{R}[g]$ denote, respectively, the corresponding Einstein tensor, Weyl tensor, Riemann tensor, Ricci tensor, and scalar curvature.

\section{Field Theories \label{secFields}}

For present purposes, we can take a \emph{classical field theory} to consist of the following elements.

\begin{enumerate}[(a)] 
\item A connected $n$-dimensional manifold $V,$ the \emph{spacetime} of the theory. 
\item A set $\Theta$ of tensors on $V.$ These are the \emph{fixed fields} of the theory. $\Theta$ is often the empty set in cases of interest.
\item A set $\{ \phi_1, \ldots, \phi_k\}$ of \emph{dynamical fields} on spacetime. A field is specified by specifying a type of tensor on $V$ and a particular configuration of the field by specifying a tensor of that type. A configuration of the complete set of dynamical fields is denoted $\Phi=(\phi_1,\ldots, \phi_k).$
\item A space $\mathcal{K}$ of field configurations, thought of as consisting of the kinematically possible $\Phi.$ In typical cases, $\mathcal{K}$ is determined by specifying the smoothness, asymptotic behaviour, etc. of the fields of the theory.
\item A set of differential equations $\Delta(\Phi; \Theta)$ that determines the space $\mathcal{S} \subset \mathcal{K}$ of \emph{solutions} of the theory. The fixed fields play the role of parameters rather than variables in $\Delta.$ Any derivative operators appearing in $\Delta$ must be definable in terms of the fixed and dynamical fields of the theory. \end{enumerate}

The \emph{symmetry group}, $\mathcal{G},$ of a classical field theory is the group consisting of diffeomorphisms from $\mathcal{K}$ to itself that map solutions to solutions and that are suitably local on $V.$\footnote{Here is one way to make this notion precise \cite{Zuckerman:1987gb}. If $G$ is a one-parameter group of diffeomorphisms from $\mathcal{K}$ to itself, then it makes sense to speak of the infinitesimal generator $\xi$ of $G.$ $\xi$ will be a vector field on $\mathcal{K}.$ In the present setting, a vector $\delta \Phi \in T_\Phi \mathcal{K},$ $\Phi \in \mathcal{K},$ can be identified locally with a tensor field on $V.$ We call $G$ \emph{local in $V$} if there is some $k$ such that for any $x \in V$ the value of $\xi (\Phi) \in T_\Phi \mathcal{K}$ when evaluated at $x$ depends only on the value of $\Phi$ and its first $k$ derivatives at $x.$ We call $G$ a \emph{one-parameter symmetry group} of the field theory if it is local and maps solutions to solutions. We take $\mathcal{G}$ to be the group generated by all of the one-parameter symmetry groups of the theory.} We call the elements of $\mathcal{G}$ the \emph{symmetries} of the theory and write $\gamma \cdot \Phi$ for the result of acting on $\Phi \in \mathcal{S}$ by a symmetry $\gamma \in \mathcal{G}.$

We will be especially interested in symmetries corresponding to diffeomorphisms of $V.$ We will denote by $\mbox{Diff}(V)$ the group of diffeomorphisms $V$ and by $\mbox{Diff}_{c}(V)$ the group of compactly supported diffeomorphisms of $V$ (so $d \in \mbox{Diff}(V)$ is in $\mbox{Diff}_{c}(V)$ if and only if there is a compact set $K \subset V$ such that $d$ is the identity on $V / K$). Since we are working with tensor fields, each field has a well-defined transformation law $\phi \mapsto d \cdot \phi$ under the action of spacetime diffeomorphisms. Further, the action of diffeomorphisms on the space of fields is local in $V.$ So the following is a special case of our general notion of symmetry. 
\begin{definition}[Spatiotemporal Symmetry] A diffeomorphism $d: V \to V$ is a \emph{spatiotemporal symmetry} of a theory if $\Phi \in  \mathcal{S}$ implies $d \cdot \Phi \in \mathcal{S}.$ \end{definition}
\begin{definition}[Local General Covariance] We call a field theory \emph{locally generally covariant} if every $d \in \mbox{Diff}_{c}(V)$ is a spatiotemporal symmetry of the theory. \end{definition}
\begin{definition}[General Covariance] We call a field theory \emph{generally covariant} if every $d \in \mbox{Diff}(V)$ is a spatiotemporal symmetry of the theory. \end{definition}

\section{Examples \label{secEx}}

In this section, a number of examples are presented as motivation for the following claims.

\begin{enumerate}[(1)] 
\item Background-independence admits of degrees, with theories falling in between full background-independence and full background-dependence.
\item A degree of background-dependence can result from the imposition of asymptotic boundary conditions.
\item The notion of background-independence is not purely formal: whether a theory is background-independent depends on how it is interpreted.
\end{enumerate}

\subsection{More or Less Background-Dependent Theories}

In paradigm background-dependent theories, solution-independent background structure is encoded in  fixed fields.

\begin{example} \label{exKG} Consider the theory of a scalar field propagating on Minkowski spacetime in which the spacetime metric $\eta$ is treated as a fixed field: spacetime is $V=\mathbb{R}^n$ ($n\geq 2$); the only fixed field is $\eta$; the only dynamical field is a scalar field $\phi$; the space $\mathcal{K}$ of kinematic possibilities is the space of twice-differentiable $\phi$; and the equation of motion is the Klein--Gordon equation, $\Box _{\eta} \phi =0.$ Here we have full background-dependence: the Minkowski metric $\eta$ is provides the fixed stage against which the scalar field evolves. 
\end{example}

Of course one can have background-dependence even in the absence of fixed fields.

\begin{example} \label{exKGgc} Consider the generally covariant analogue of the theory of the preceding example: Let $V=\mathbb{R}^n$ ($n\geq 2$); let there be no fixed field; let there be two dynamic fields, a scalar field $\phi$ and a Lorentz-signature metric $g$; the space $\mathcal{K}$ of kinematic possibilities is the space of twice-differentiable $\phi$ and $g$; and the equations of motion are $\mathbf{Riem}[g] = 0$ and  $\Box_{g}\phi = 0.$ Each solution $(g, \phi)$ just represents a Klein--Gordon field on a copy of Minkowski spacetime: the physics is exactly the same as in the preceding example. But whereas in the preceding example the group of spatiotemporal symmetries was given by the Poincar\'e group, here it is the group $\mbox{Diff}(V)$ of spacetime diffeomorphisms. Despite its general covariance, this theory is background-dependent: although the metric $g$ is a dynamical field, the spacetime geometry is solution-independent and plays the role of stage against which the scalar field evolves. 
\end{example}

It possible to cook up variants on this example that, intuitively, fall just short of full background-dependence. 

\begin{example} \label{exTorus} Let everything be as in the preceding example, except that $V$ is taken to have the topology of $ \mathbb{R} \times \mathbb{T}^{m},$ where $\mathbb{T}^{m}$ is the $m$-dimensional torus and it is built into the definition of $\mathcal{K}$ that $g$ is globally hyperbolic.

Once again, the metric $g$ is flat in any solution. But the theory has nontrivial global geometric degrees of freedom---the space of metrics modulo diffeomorphisms is finite-dimensional.\footnote{For relevant degrees of freedom in the case of three spacetime dimensions, see, e.g., \cite{Carlip:1998zo,Louko:1994yg}%Carlip[\S 3.3]
. For their operational significance, see \cite{Meusburger:2009yq}.} 

This theory is of course about as far from being background-independent as is possible---at the local level, the spacetime geometry is Minkowskian in every solution, and it is this geometry that determines the behaviour of the scalar field. But because the global geometry varies from solution to solution, it is natural to see the theory as falling (just) short of the full background dependence of the previous example.  \end{example}

The next example is a variation on this theme which raises a point that will play a role the discussion below. 

\begin{example} \label{exDS} 
Let $V\simeq  \mathbb{R} \times S^{3}$ ($S^{3}$ being the three-sphere). We again have no fixed fields and two dynamic fields, a scalar field $\phi$ and a metric $g.$ The space of kinematic possibilities is determined by the usual differentiability conditions together with the condition that $g$ be geodesically complete. The equations of motion are 
\begin{eqnarray*}  \mathbf{Weyl}[g] &  = &  0 \label{eqnWeyl} \\ \mathbf{Ein}[g] + \frac{g}{4}\mathbf{R}[g] &  = &  0 \\ \Box_{g}\phi & = & 0. \label{eqnDS} \end{eqnarray*}
This theory describes a scalar field propagating against a de Sitter spacetime of constant curvature $K^2$: if $(g,\phi)$ is a solution, then $(V,g)$ is isometric to a timelike hyperboloid of radius $K$ in Minkowski spacetime.\footnote{The equations of motion force $(V,g)$ to be a space of constant curvature $K^2$ \cite[p. 124]{Hawking:1973gk}. The assumption that $g$ is geodesically complete then forces $(V,g)$ to be a quotient of anti-de Sitter spacetime, Minkowski spacetime, or de Sitter spacetime \cite[Theorem 2.4.9]{Wolf:1984yq}. But the topology of $V$ rules out the first two options and forces $(V,g)$ to be de Sitter spacetime itself.}

The equations of motion almost succeed in fixing the spacetime geometry up to diffeomorphism---in this theory, identifying metrics related by a diffeomorphism leaves us with a single geometric degree of freedom, parameterized by $K^2.$ If we regard scale transformations as well as diffeomorphisms as relating physically equivalent solutions, then we should regard this theory as fully background-dependent---for then there are no geometrical degrees of freedom on the theory. However, if we regard scale transformations as physical, then we should regard this theory as nearly but not entirely background-dependent in virtue of possessing a single geometrical degree of freedom. \end{example} 

\subsection{More or Less Background-Independent Theories}

At the other end of the spectrum from the examples we have been considering lies spatially compact vacuum general relativity, a paragon of background-independence. 

\begin{example} \label{exGR} Let $V$ be an $n$-manifold of the form $\mathbb{R} \times \Sigma$ where $\Sigma$ is compact and let the only field be a twice-differentiable globally hyperbolic Lorentzian metric $g$  subject to the vacuum Einstein equation, $\mathbf{Ricci}[g]=0.$ This theory is generally covariant. It is also background-independent: the geometry of spacetime is determined dynamically rather than being imposed \emph{ab initio}.\footnote{Of course, this is not to say that say that there is \emph{nothing} that is invariant across solutions: each solution involves a Lorentz-signature metric, each solution satisfies the field equation of the theory, etc.}  \end{example}

Just as there are theories that fall just short of background-dependence, there are also theories that fall just short of background-independence. 

\begin{example} \label{exGRaFlat} Consider the sector of general relativity in which solutions are asymptotically flat at spatial infinity in the sense of \cite{Andersson:1987kf}. Here $V=\mathbb{R}^{4}$ and the only field is a Lorentzian metric $g$ subject to the vacuum Einstein equation that is required to be twice-differentiable, to be globally hyperbolic, and to approximate Euclidean geometry in a suitable sense at spatial infinity.  

A helpful, but dispensable, way to think about these asymptotic boundary conditions appeals to an ideal boundary added to $V$ that represents spatial infinity. In broadest outline, the picture is as follows. We attach a non-physical boundary $\mathcal{H}$ to $V$ that has the topology and metric of $\mathcal{H}_0$ the unit timelike hyperboloid in Minkowski spacetime; for convenience we fix an isometry $d: \mathcal{H} \to \mathcal{H}_0.$ Let $g$ be a Lorentzian metric on $V$ and let $\Sigma$ be a Cauchy surface relative to $g.$ We say that $\Sigma$ has \emph{good asymptotic behaviour} relative to $g$ if the initial data that $g$ induces on $\Sigma$ represents space as asymptotically Euclidean (and satisfies various additional technical conditions) and if the set of points $\partial \Sigma \subset \mathcal{H},$ which we call \emph{the boundary of} $\Sigma,$ consisting of the limit points of $\Sigma$ on $\mathcal{H},$ has the same form as a set on $\mathcal{H}_0$ that arises by taking the intersection of $\mathcal{H}_0$ with an spacelike hyperplane (in the standard model of Minkowksi spacetime).  A metric $g$ on $V$ is in $\mathcal{K}$ if by its lights it is possible to partition $V$ by Cauchy surfaces with good asymptotic behaviour. 
  
The resulting theory is locally generally covariant---since locally its dynamics is given by the Einstein equation. But is is not generally covariant: only certain diffeomorphisms from $V$ to itself preserve the asymptotic boundary conditions. One can picture this as follows. Let $\Theta_0$ be the set consisting of all the intersections of $\mathcal{H}_0$ with spacelike hyperplanes in Minkowski spacetime. And let $\Theta := d^{-1} \Theta_0$ be the corresponding set of subsets of $\mathcal{H}.$ Any diffeomorphism $f: V \to V$ induces a diffeomorphism $\bar{f} : \mathcal{H} \to \mathcal{H}.$ $f$ is a symmetry of our theory if and only if under the obvious action $\bar{f}$  maps $\Theta$ to itself.

The most natural thing to say is that this theory lies between paradigmatic non-background-independent theories like those in in which fields propagate against the backdrop of Minkowski spacetime and paradigmatic background-independent theories like spatially compact general relativity. On the one hand, there are no fields on spacetime, fixed or dynamical, that encode a fixed background structure such as a geometry---indeed, locally the field of the theory has all of the freedom of the metric field of ordinary spatially compact general relativity. On the other, there is also a sense in which the boundary conditions of the theory ensure that any solution has the structure of Minkowski spacetime at infinity---and this is reflected in the fact that the theory is not generally covariant. Perhaps the point is put most vividly by saying that in this theory one has a class of preferred frames at infinity. An observation along these lines played an important role in motivating Einstein to investigate spatially compact cosmologies \cite{Janssen:2008rt}.%[\S 5]
 \end{example}

\subsection{Geometrically Ambiguous Theories}

In each of the above examples, there was among the fields of the theory an obvious candidate to represent the geometry of spacetime---and the question of the extent to which the theory was background-independent turned on the question of the extent to which the behaviour of that field was solution-independent. But it is also possible to find theories featuring more than one field that could plausibly represent the geometry of spacetime. In such cases the question of background-independence becomes more subtle. 

\begin{example} \label{exNord} Consider a theory modelled on Nordstr\"{o}m's scalar theory of gravity.\footnote{For discussions of Nordstr\"{o}m's theory, see \cite{Giulini:qm,Ravndal:2004yy,Straumann:2000gy}%[\S 2.4.1] %[\S 2] %[\S 2.1]
. The theory discussed here is not quite Nordstr\"{o}m's, since in his theory the scalar field coupled to the trace of the stress-energy tensor of particulate matter.} Let $V\simeq \mathbb{R}^4.$ Let there be no fixed field and let there be three dynamical fields---two metrics of Lorentz signature, $\eta$ and $g,$ and a positive scalar field $\phi,$ subject to the usual sort of smoothness conditions. In order to state the equations of motion, we also introduce a scalar field $T$ that arises by taking the trace relative to $g$ of the ordinary scalar field stress-energy tensor for $phi$ relative to $g.$ The equations of the theory fall into three groups. The first group,
\begin{subequations} \label{eqnNord1}
\begin{eqnarray}
\mathbf{Riem}[\eta] & = & 0 \label{eqnNord1a}
\\ \Box_{\eta}\phi & = & -4\pi G \phi ^{3} T, \label{eqnNord1b}
\end{eqnarray}
\end{subequations}
says that $\eta$ is a flat metric and that $\phi$ is a modified Klein--Gordon field with respect to $\eta$ with an interesting source term (here and below $G$ denotes Newton's constant). The second group,
\begin{subequations} \label{eqnNord2}
\begin{eqnarray}
\mathbf{Weyl}[g] & = & 0 \label{eqnNord2a}
\\ \mathbf{R}[g] & = & 24\pi G T, \label{eqnNord2b}
\end{eqnarray}
\end{subequations}
says that $g$ is conformally flat and has scalar curvature proportional to $T.$ The third group,
\begin{subequations} \label{eqnNord3}
\begin{eqnarray}
g & = & \phi^{2} \eta \label{eqnNord3a}
\\ \phi & = & (- \mbox{det} g )^{\frac{1}{8}} \label{eqnNord3b}
\\ \eta & = & g (- \mbox{det} g )^{-\frac{1}{4}}, \label{eqnNord3c}
\end{eqnarray}
\end{subequations}
encodes relations between $g$ on the one hand and $\phi$ and $\eta$ on the other. These three sets of equations are highly redundant: in the presence of the third group, the first and second groups are equivalent to one another.  

Is this theory background-independent? This depends upon how we understand the theory.\footnote{For discussion of options for interpreting such theories, see \cite[\S\S 3.III and 3.IV]{Sklar:1985wk}. Note that if one were to couple this theory to particulate matter in the obvious way (as Norstr\"{o}m in fact did) then physical considerations would exert some pressure in favour of the second option discussed below \cite{Norton:1992eb}%[\S\S 12--14]
.} 

On the one hand, one could take $\phi$ and $\eta$ to be the fundamental physical fields of the theory and understand the theory as describing the propagation of a non-linear scalar field in Minkowski spacetime ($g$ would then just be a clever but unphysical way of encoding $\phi$ and $\eta$ in a single object obeying an elegant equation). Understood this way, the theory is as fully background-dependent as the theory of a Klein--Gordon field in Minkowski spacetime. 

Alternatively, we could think of $g$ as the fundamental physical variable, understood as directly representing the geometry of spacetime. In this case we would view Equations (\ref{eqnNord2a}) and (\ref{eqnNord2b}) as giving the laws of a non-linear field theory of spacetime geometry (Equations (\ref{eqnNord3b}), (\ref{eqnNord3c}), and (\ref{eqnNord1b}) would then tell us how to rewrite the theory as a theory of a scalar field propagating against a non-physical flat metric). In this case we would presumably want to consider the theory as being far from background-dependent: just as in general relativity, here we have a non-linear theory of a field with infinitely many degrees of freedom describing a spacetime geometry that is in general variably curved.   \end{example}

\subsection{Morals}
\begin{enumerate}[(1)]
\item Background-dependence and independence come in degrees: some theories are fully background-(in)dependent, others only nearly so---and others fall somewhere in between. 
\item A theory can fail to be fully background-independent in virtue of asymptotic boundary conditions.
\item The extent of the background-(in)dependence of a theory is not a strictly formal one: in particular, it depends on how one thinks of the geometric structure of each solution and on what sorts differences between solutions one takes to be unphysical. \end{enumerate}

\noindent Running through the discussion of the above examples was the idea that a theory is fully background-independent if each physical possibility corresponds to a distinct spacetime geometry, and that it falls short of full background-independence to the extent that this condition fails. The task of the next several sections will be to put in place a framework of concepts that will allow us to make this idea precise. 

\begin{remark}[Absolute Objects] \label{remAbs} General relativity was the first physical theory in which space and time did not have the objectionable feature of acting on matter without being acted upon by it.\footnote{This was a favourite theme of Einstein's; see  \cite[\S 3.9]{Norton:1993fv}.} The notion of an object that acts upon others without itself being acted upon is given a precise sense in the notion of an of an \emph{absolute object} due to Anderson and Friedman: an absolute object of a theory is one that is locally the same up to diffeomorphism in every solution \cite{Anderson:1964nr,Anderson:1967tu,Earman:1974te,Friedman:1973nw,Friedman:1983rc}.\footnote{This notion has its problems---see \cite{Pitts:2006lk} for a survey. Foremost among them is the fact that there are certain types of tensors, such as non-vanishing vector fields or symplectic forms, that are always locally identical up to diffeomorphism---so any field represented by such an object counts as absolute, no matter what its dynamics.} %\cite[p. 59 n. 9]{Friedman:1983rc}

Since an absolute object of a theory is a field that is \emph{the same} in each solution, it may seem natural to take a theory to be background-independent if and only if it features no absolute objects. But this suggestion founders on the morals reached above. For if one identifies background-independence with the lack of an absolute object then: (i) background-independence will be an all or nothing affair; (ii)  if a theory fails to be background-independent, this is always in virtue of a the fact that some field of the theory is an absolute object; and (iii)  the question of background-independence is rendered purely formal.  \end{remark}

\section{Geometric Degrees of Freedom} \label{secGeom}

Suppose that one is handed a field theory (in the sense of Section \ref{secFields} above). So far it is just a piece of mathematics. In order to endow it with physical content, one would have to say something about how the fields of the theory correlate with observable quantities. In particular, if the field theory is to be understood as an all-encompassing account of the classical world, then it must be endowed with geometric content: one must have some way of thinking of each solution of the theory as portraying physical processes in spacetime.

Let us say that a \emph{geometrization} for a field theory consists of: (i) a rule $\Phi \mapsto \mathfrak{g}_\Phi$ that associates with each solution $\Phi$ of the theory a geometric structure $\mathfrak{g}_\Phi$ for the manifold $V$ on which the fields of the theory live; and (ii) a criterion that tells us when geometries $\mathfrak{g}_{\Phi_1}$ and $\mathfrak{g}_{\Phi_2}$ are geometrically equivalent. 

When we think of a field theory as a relativistic field theory, we are implicitly thinking of it as endowed with a geometrization that: (i) assigns to each solution $\Phi$ a Lorentzian metric $\mathfrak{g}_\Phi$ on $V$ that is among the fields of the theory or definable in terms of them; and (ii) counts such metrics as geometrically equivalent if and only they are related to one another by a diffeomorphism from $V$ to itself. 

There are of course other notions of geometrization that arise. Pre-relativistic field theories can be thought of as involving a geometrization that: (i) assigns to each solution a spatial metric, a temporal metric, and an affine connection; and (ii) counts two such assignments as geometrically equivalent if and only they are related by a diffeomorphism. One might also tinker with the standard notion of a relativistic field theory by counting metrics related by scale transformations as geometrically equivalent.

In most of the examples discussed in the preceding section, a default reading of the theory as a relativistic field theory was available, since one and only one of the fields of the theory was a Lorentzian  metric. In the final example discussed there were two fields of this type---and the question arose which one should be taken to describe the spatiotemporal geometry of solutions. Other situations are of course possible. A field theory may admit a natural reading as a relativistic field theory even though there are no Lorentzian metrics among its fields.\footnote{Think here of the formulation of $(2+1)$ gravity as a Chern--Simons theory \cite{Carlip:1998zo}%[Chapters 2 and 4]
. or of certain `parameterized' versions of familiar theories \cite{Gotay:2010yq,Lee:1990pm,Torre:1992ze}.} Or a theory that includes one or more Lorentzian metrics among its fields may admit a natural geometrization by whose lights the geometry of a solution is given by some further metric (Nordstr\"om's theory could be put in this form, as could the Brans--Dicke theory).

Relative to a geometrization for a theory, we can look at the set of spacetime geometries that arise in solutions of the theory and at the set of equivalence classes of such geometries under the relation of geometrical equivalence of the geometrization. We call any set of variables that parameterize this latter set the \emph{geometrical degrees of freedom} of the theory relative to the geometrization. 

Relative to the standard notion of geometrization, the theories of Examples  \ref{exKG}
and \ref{exKGgc} above have trivial geometrical degrees of freedom (up to equivalence, only one geometry occurs); the theories of Examples  \ref{exTorus} and \ref{exDS} have finitely many geometrical degrees of freedom (the space of geometries modulo equivalence is finite-dimensional); and the theories of Examples \ref{exGR} and
\ref{exGRaFlat} have infinitely many geometrical degrees of freedom (the space of geometries modulo equivalence is the infinite-dimensional space of Ricci-flat Lorentzian metrics modulo diffeomorphisms).\footnote{Relative to the unorthodox approach that counts metrics related by scale transformations as equivalent, the theory of Section \ref{exDS} has trivial geometrical degrees of freedom.} Example \ref{exNord}, Nordstr\"om's theory, is more subtle: relative to one way of reading the theory, it has trivial geometrical degrees of freedom; relative to another, it has infinitely many.

\section{Physical Degrees of Freedom} \label{secCount}

This section focusses on the way of counting physical degrees of freedom that is standard among physicists (an alternative system of counting, more popular among philosophers than among physicists, is mentioned in Remark \ref{remSmolin} below). 

In many cases, one takes the physical degrees of freedom of a theory to parameterize the space of solutions.\footnote{Actually this is non-standard---the space of solutions is isomorphic to the system's phase space, and the degrees of freedom are usually taken to parameterize the system's configuration space. But it is convenient here to omit the factor of one-half that should be inserted at various points below.} But there are important exceptions. 

The equations of motion of a classical theory are said to be \emph{underdetermined} if they are not independent of one another; for a precise characterization, see \cite[p. 171]{Olver:1993qp}.  The most prominent types of theories with underdetermined equations include generally covariant theories and theories of Yang--Mills type. But there are many other examples \cite{Henneaux:1992yp}. Uniqueness of solutions fails radically in theories with underdetermined equations of motion: the family of solutions corresponding to an admissible set of initial data is infinite-dimensional---roughly speaking, such a family can be parameterized by arbitrary functions of the independent variables of the theory.

Since a theory is deterministic if and only if each instantaneous state is compatible with only one global history, there is usually a tight connection between a failure of uniqueness of solutions in a theory and the failure of that theory to be deterministic. But this connection obtains only if we assume that distinct solutions of our theory always represent physically distinct situations. Faced with the prospect of a wholesale and dramatic failure of determinism in the presence of underdetermined equations of motion, one usually prefers to reject this assumption. Standardly one assumes instead that a theory featuring underdetermined equations involves  
 gauge freedom---one assumes, that is, that some of the degrees of freedom of the theory are unphysical and that, except perhaps in special cases, specifying instantaneous values of all variables suffices to determine the past and future behaviour of the physical degrees of freedom.
 
The idea is to introduce introduce an equivalence relation---\emph{gauge equivalence}---on the space of solutions, then to identify the physical degrees of freedom of the theory with variables that parameterize the quotient space that results when we identify gauge equivalent solutions. 

How can we characterize gauge equivalence? Roughly speaking, we want to consider solutions to be gauge equivalent if the underdetermination of the theory's equation of motion forces us to consider them physically equivalent, on pain of taking the theory to be indeterministic. But we cannot just take two solutions to be gauge equivalent if and only if they induce the same initial data at some instant of time. 

\begin{enumerate}[(i)] \item From the fact that solutions $\Phi_1$ and $\Phi_2$ induce the same initial data at time $t_1$ and the fact that solutions $\Phi_2$ and $\Phi_3$ induce the same initial data at time $t_2,$ it need not follow that there is a time at which $\Phi_1$ and $\Phi_3$ induce the same initial data. But we need gauge equivalence to be an equivalence relation, so we need to set things up so that from the fact that $\Phi_1$ and $\Phi_2$ are gauge equivalent and the fact that $\Phi_2$ and $\Phi_3$ are gauge equivalent, it follows that $\Phi_1$ and $\Phi_3$ are gauge equivalent.
\item Simply identifying solutions whenever they induce the same initial data will efface genuine instances of indeterminism. The Newtonian $n$-body problem admits solutions in which a system of particles interacts in such a way that they all disperse to spatial infinity in finite time \cite{Diacu:1992cj}. Such a solution matches the trivial empty solution at late times. This constitutes real physical indeterminism, not an instance of gauge equivalence. Similarly, in the case of spatially compact general relativity, we will want to consider solutions gauge equivalent if and only if they are related by a diffeomorphism. Instances in which globally hyperbolic solutions admit non-isometric extensions are instances of genuine indeterminism, not gauge equivalence; for examples see \cite{Chrusciel:1993ol}.
\end{enumerate}

There are a number of ways of addressing the problem of characterizing gauge equivalence. For theories in Lagrangian form, one can employ by-products of the variational procedure to construct a presymplectic form on the space of solutions of the theory and take two solutions to be gauge equivalent if and only if they can be connected by a (piecewise) null curve of this presymplectic form \cite{Crnkovic:1987ju,Deligne:1999tu,Zuckerman:1987gb}. Alternatively, one can follow the Dirac constraint algorithm, which leads from a Lagrangian to a notion of gauge equivalence on the space of initial data of the Hamiltonian formulation of the theory \cite{Dirac:2001el,Gotay:1978li,Henneaux:1992yp}. These two approaches are closely related to one another \cite{Lee:1990pm}. 

There is also a low-tech procedure that gives the same answers in standard applications. Let us take for granted that we know, for any solution $\Phi,$ which hypersurfaces in $V$ correspond to initial data surfaces (=instants of time). Then we can make the following definitions. Let us say that a symmetry $\gamma$ of a theory is a \emph{spoiler} relative to solution $\Phi$ if there is some initial data surface $\Sigma \subset V$ relative to $\Phi$ such that $\gamma$ acts as the identity on some neighbourhood $U$ of $\Sigma.$\footnote{I.e., for any $x \in U,$ $(\gamma \cdot \Phi)(x)=\Phi(x).$ In vacuum general relativity, the spoilers are the diffeomorphisms that act as the identity on a neighbourhood of a Cauchy surface; in Maxwell's theory set in Minkowski spacetime (with a fixed metric) the spoilers are the transformations of the form $A \mapsto A +d\Lambda$ where $\Lambda$ is a scalar function that vanishes on a neighbourhood of a hyperplane of simultaneity.} We call solutions $\Phi_1$ and $\Phi_2$ \emph{spoiler-related} if there is a spoiler $\gamma$ relative to $\Phi_1$ such that $\Phi_2=\gamma \cdot \Phi_1.$ Being spoiler-related is not an equivalence relation (for the sort of reasons raised in point (i) above). We call an equivalence relation $R$ an \emph{extension} of the relation of being spoiler-related if any two spoiler-related solutions are also $R$-related. There are many equivalence relations between solutions that extend the relation of being spoiler-related. One such is the relation that takes any two solutions to be equivalent. That relation is maximally strong. We are after a much weaker one (i.e., one according to which fewer pairs of solutions are equivalent). Heuristically, what we want is the weakest equivalence relation that extends the spoiler-relatedness.\footnote{This the the relation $R$ that takes solutions $\Phi$ and $\Phi^\prime$ to be equivalent if and only if there are solutions $\Phi_1,$ \ldots, $\Phi_n$ such that $\Phi_1=\Phi,$ $\Phi_n=\Phi^\prime,$ and each $\Phi_i$ is spoiler-related to $\Phi_{i+1}$ ($i=1,\ldots, n-1$).} For technical convenience, we take \emph{gauge equivalence} to be the weakest equivalence relation on the space of solutions that extends spoiler-relatedness and which has equivalence classes that form submanifolds of the space of solutions---if there is any such relation.

One can show (modulo certain technical questions) that for various sectors of general relativity this definition is well-formed and gives the same answer as the high-tech Lagrangian approach \cite{Belot:2008yq}. In particular for spatially compact general relativity, both approaches agree that solutions are gauge equivalent if and only if they are related by diffeomorphisms. And for well-behaved asymptotic boundary conditions at spacelike, timelike, or null infinity, both approaches agree that solutions are gauge equivalent if and only if they are related by diffeomorphisms that act as the identity at the appropriate asymptotic boundary.\footnote{In some cases, this requires taking initial data surfaces to correspond to (partial) Cauchy surfaces with good asymptotic behaviour.}  

To recap. Whichever way one handles the details, the idea is that we take solutions to be gauge equivalent if the formalism of the theory forces us to consider them physically equivalent on pain of recognizing rampant and uninteresting indeterminism. We then identify the space of physically distinct possibilities described by the theory with the space of solutions modulo gauge equivalence, and identify the physical degrees of freedom of the theory with any set of variables that parameterizes that quotient space.

\begin{remark}[Non-Standard Counting] \label{remSmolin} On the standard approach discussed above, symmetries that relate gauge-equivalent solutions have a quite different status from other symmetries: indeed, the idea of gauge equivalence is often introduced by saying that whereas global symmetries like Lorentz symmetries relate physically distinct situations, the symmetries characteristic of Yang--Mills theories are non-physical and their presence an indication of redundancy in the variables of the theory. 

There is an alternative approach on which all symmetries are on equal footing: in order to find the degrees of freedom of a theory, one has to quotient out all symmetries---including those that are not gauge in the present sense. Thus, in the case of a theory whose equations of motion are not under-determined, one must quotient the space of solutions by the spatiotemporal and other symmetries of the theory in order to arrive at the true degrees of freedom. For this approach, see \cite{Smolin:2001rs,Smolin:2006yq}. 
\end{remark}

\section{Background-Independence} \label{secBI}
 
In paradigm background-dependent theories, the geometry is the same from solution solution to solution: vary the physical degrees of freedom as you will, you cannot alter the geometry of spacetime. What is the very opposite of such behaviour? A theory in which \emph{any} change in the physical degrees of freedom makes a difference to the geometry of spacetime. This motivates the following definitions.
\begin{definition}[Full Background-Dependence] A field theory is \emph{fully background-dependent} if it has no geometrical degrees of freedom: every solution is assigned the same spacetime geometry as every other solution. \end{definition}
\begin{definition}[Full Background-Independence] A field theory is \emph{fully background-independent} if all of its physical degrees of freedom correspond to geometrical degrees of freedom: two solutions correspond to the same physical geometry if and only if they are gauge equivalent. \end{definition}
Note that both of these definitions presuppose that a notion of geometrization is in place; the second also requires a notion of gauge equivalence. 

In effect, these definitions measure the degree of background-independence by looking at how many of the physical degrees of freedom correspond to geometrical degrees of freedom. The limiting cases are full background-dependence (in which there are no geometrical degrees of freedom) and full background-independence (in which all physical degrees of freedom are geometrical degrees of freedom). Intermediate cases are also possible.

\begin{definition}[Near Background-Dependence] A field theory is \emph{nearly background-dependent} if it has only finitely many geometrical degrees of freedom: quotienting the space of geometries that arise in solutions of the theory by the relation of geometrical equivalence yields a finite-dimensional space. \end{definition}
\begin{definition}[Near Background-Independence] A field theory is \emph{nearly background-independent} if it has a finite number of non-geometrical degrees of freedom: there is some $N$ such that for any geometry arising in a solution of the theory, the space of gauge equivalence classes of solutions with that geometry is no more than $N$-dimensional.  \end{definition}

These definitions give the desired verdicts concerning the examples of Section \ref{secEx} above.

\begin{enumerate}[ ]
\item The theories of Examples \ref{exKG} and \ref{exKGgc} describe a Klein--Gordon field in Minkowski spacetime (with or without  the metric as a fixed field). Under the natural geometrization, these theories are fully background-dependent: each solution has the geometry of Minkowski spacetime, so there are no geometrical degrees of freedom.\footnote{Here and below, the natural geometrization of a theory treats the theory as a relativistic field theory in the sense of Section \ref{secGeom} above, with the geometry of each solution given by the metric tensor that appears as one of the fields of the theory.} 
\item The theory of Example \ref{exTorus} concerns a scalar field propagating against a flat but dynamical metric on a spacetime with toroidal spatial topology. Under the natural geometrization it is nearly background-dependent: each solution has the local geometry of Minkowski spacetime, but a finite number of global geometrical degrees of freedom remain.   
\item The theory of Example \ref{exDS} describes a scalar field propagating against a de Sitter metric. Under the natural geometrization, the theory is nearly background-dependent: the theory has a single geometrical degree of freedom (the value of the curvature constant). Note, however, that the theory would count as fully background-dependent if one chose to count metric tensors related by scaling transformations as geometrically equivalent.
\item The theory of Example \ref{exGR} is spatially compact vacuum general relativity. Under the natural geometrization and notion of gauge equivalence, the theory is of course fully background-independent: two solutions share the same geometry if and only if they are related by a diffeomorphism if and only if they are gauge equivalent.
\item The the theory of Example \ref{exGRaFlat} is vacuum general relativity with asymptotic flatness imposed at spatial infinity. Under the natural geometrization and the standard notion of gauge equivalence, the theory is nearly background-independent: two solutions correspond to the same geometry if and only they are related by a diffeomorphism; two solutions are gauge equivalent if and only if they are related by a diffeomorphism asymptotic to the identity at spatial infinity; if one quotients the family of solutions sharing a given geometry by the relation of gauge equivalence, the result is a ten-dimensional space.\footnote{The quotient of the group of diffeomorphisms by the group of diffeomorphisms acting as the identity at spatial infinity is the Poincar\'e group. This group acts as a symmetry group on the space of gauge equivalence classes. See 
\cite{Andersson:1987kf}.} Note, however, that if one were to adopt the non-standard method of counting physical degrees of freedom discussed in Remark \ref{remSmolin} above, then this theory would count as fully background-independent.
\item The theory of Example \ref{exNord} is inspired by Nordstr\"om's bimetric theory of gravity: its fields include a metric tensor $\eta$ that is flat in every solution and a metric tensor $g$ that varies from solution to solution. If the first of these fields is taken to give the spacetime geometry of the theory, then the theory is fully background-dependent (spacetime is $\mathbb{R}^4,$ so there are no global degrees of freedom). If instead the metric $g$ is taken to give the spacetime geometry, then the theory is not background-dependent, nor even nearly background-dependent: there are infinitely many geometrical degrees of freedom. 
\end{enumerate}

\section{Discussion} \label{secDisc}

The account of background-(in)dependence advanced above embodies the three morals developed in Section \ref{secEx}. (i) Background dependence and independence come in degrees. (ii) Background-independence can be spoiled by imposition of asymptotic boundary conditions, as well as by the presence of solution-invariant fields in the theory. (iii) The status of a theory \emph{vis-\`a-vis} background-dependence and independence can depend on the interpretation given to the theory as well as on its formalism. 

In this final section, a number of loose ends and further topics are taken up: the possibility of translating the framework into a setting that allows more general sorts of field theories; further niceties concerning the dependence of background-(in)dependence on how theories are understood; further points about asymptotic boundary conditions; and how adding matter to general relativity affects the question of background-independence.

\subsection{More General Field Theories}

So far, a relatively narrow notion of a classical field theory has been employed, according to which all fields are tensor fields. More generally, one might allow configurations of fields to be given by sections of arbitrary fibre bundles over spacetime. The notions of geometrization, spoilers, and gauge equivalence of Sections \ref{secGeom} and \ref{secCount} above carry over to this more general setting---and so the notions of background-(in)dependence given in Section \ref{secBI} continue make sense. Whether these definitions still lead to reasonable results is a question for further investigation. Of course, not all fields have a well-defined transformation law under diffeomorphisms. So the notion of general covariance only makes sense for certain types of fields---those fields whose configurations are sections of a natural bundle over spacetime (see \cite{Kolavr:1993yq}).

\subsection{Dependence on Interpretation}

We have already seen that the notions of background-dependence, background-independence, etc., are not strictly formal notions: the status of given theory may depend on how the formalism of a theory is interpreted. This fact is reflected in the definitions given in Section \ref{secBI} above: the degree of background-(in)dependence of a theory may depend on which of the fields of a theory are taken to represent geometrical structure, on the standards used to determine whether two geometrical structures are `the same,' and on whether one views solutions represented by (non-gauge) symmetries as always corresponding to the same physical situation.\footnote{For illustrations, see the discussions in the preceding section of Nordstr\"om's theory, of the de Sitter-based theory, and of general relativity with asymptotic flatness imposed at spatial infinity.}

Here is another illustration. In setting up a model of an elastic continuum (a beam, a shell, a string \ldots) in a fixed spacetime background one can work with so-called \emph{material variables} \cite{Marsden:1994rt}: one begins with a space $B$ whose points correspond to the material points of the continuum and identifies a history of the continuum with a map $\phi$ from $\mathcal{B}:=B \times \mathbb{R}$ to a Lorentzian manifold $(V_0,g_0)$ (with $\phi$ a diffeomorphism onto its image and the image of each $\{b\}\times \mathbb{R},$ $b\in B,$ a timelike curve). Such a theory can be considered a classical field theory in more or less the sense of Section \ref{secFields} above, with  $\mathcal{B}$ being the `spacetime,' and the dynamical field of the theory being the $V_0$-valued field $\phi.$ There is an obvious way to geometrize such a theory by viewing each solution $\phi$ as endowing $\mathcal{B}$ with the structure of a relativistic spacetime---just pull $g_0$ back to $\mathcal{B}$ via $\phi.$\footnote{The Nambu--Goto action for string theory takes $\phi^*g$ to be the fundamental dynamical variable. One can do something similar in continuum mechanics \cite{Cairlet:2009yq}.} If one now applies the definitions of Section \ref{secBI}, one typically finds that so-considered the theory is far from being background-dependent---in general distinct embeddings $\phi$ and $\phi^\prime$ lead to non-isometric pull-backed metrics on $\mathcal{B}$  (unless boundary conditions have been imposed that fix the shape of the boundary of $\phi({B})$). But of course this is all based on a \emph{misreading} of the theory. The physical spacetime of the theory is $V_0,$ not $\mathcal{B}$---and the theory counts as fully background-dependent when reformulated of as a classical field theory with $V_0$ as its spacetime and with physical geometry given by the fixed field $g_0.$

\subsection{Asymptotic Boundary Conditions} 

An important feature of the apparatus of Section \ref{secBI} is that it allows one to isolate the sense in which the imposition of asymptotic boundary conditions can spoil background-independence. In the case of general relativity, the sorts of asymptotic boundary conditions usually discussed can be thought of in the following terms. One adjoins to the $n$-dimensional physical spacetime manifold $V$ a non-physical $(n-1)$-dimensional boundary $\partial V$ endowed with some sort of geometrical structure. Kinematically possible $\Phi$ are required to exhibit good behaviour as they approach $\partial V$; likewise  initial data surfaces should be asymptotic to subsets of $\partial V$ that are geometrically nice. We denote by $\mathcal{D}$ the group of spatiotemporal symmetries of our theory and by $\mathcal{D}_0$ the group of spatiotemporal symmetries asymptotic to the identity at $\partial V.$ In typical examples $\mathcal{D}_0$ is a proper subgroup of $\mathcal{D}$ which is a proper subgroup of $\mbox{Diff}(V).$ Further, one typically finds that $\mathcal{D}_0$ is a normal subgroup of $\mathcal{D}$ and that the quotient group $G=\mathcal{D}/\mathcal{D}_0$ can be thought of as acting geometrically at $\partial V$ (i.e., every diffeomorphism in $\mathcal{D}$ induces an isomorphism from $\partial V$ to itself, with diffeomorphisms that differ by an element of $\mathcal{D}_0$ inducing the same isomorphism). When this obtains, $G$ is called the \emph{asymptotic symmetry group.} 

For the most common sorts of asymptotic boundary conditions for vacuum general relativity, one knows or expects that (relative to a suitable notion of initial data surface) solutions are gauge equivalent if and only if they are related by a diffeomorphism asymptotic to the identity at $\partial V$---and hence that (when all goes well) the asymptotic symmetry group acts on the space of gauge equivalence classes of solutions  
(for several examples and one counter-example, see \cite{Belot:2008yq}).

If all of this obtains, then under its natural reading the theory is nearly background-independent so long the asymptotic symmetry group $G$ is finite-dimensional---as it was in Example \ref{exGRaFlat} above (where $G$ was the Poincar\'e group). But even when this picture holds, strange things can happen: when one imposes asymptotic flatness at null infinity, the asymptotic symmetry group is the infinite-dimensional  Bondi--Metzner--Sachs group (see, e.g., \cite{Geroch:1977vy}); and one expects that this group acts on the space of gauge equivalence classes of solutions \cite[example 4]{Belot:2008yq}. This means that the theory comes out as not nearly background-independent: even after the geometry of spacetime is fixed, an infinite number of physical degrees of freedom remain free. This is an unsettling conclusion---but perhaps it can be accepted as being of a piece with the strangeness of the symmetry group of this sector of general relativity.\footnote{It is sometimes suggested that the complicated asymptotic structure of this sector is a reflection of the fact that radiation can reach conformal infinity in this setting \cite{Anninos:yq}.}

One might also worry about a problem at the other end of the spectrum. What if one were to impose \emph{asymmetric} asymptotic boundary conditions? E.g., what if one were to modify the standard framework for asymptotically anti-de Sitter spacetimes in such a way that the asymptotic symmetry group became trivial \cite{Ashtekar:1984vw}%[p. L42]
? According to the present approach, the resulting theory would be fully background-independent, despite the presence of asymptotic boundary conditions! Presumably, however, this sort of theory is ruled out by our decision to focus on theories of the entire universe---it is difficult to see how the need for asymmetric asymptotic boundary conditions could arise for theories of this type.

\subsection{General Relativity with Matter} \label{ssecMatter}

So far, in considering general relativity, we have limited attention to the vacuum case. What happens when matter is introduced? That depends. 

Consider first what happens if one includes a scalar field $\phi$ that sees but is not seen by the metric $g$ (i.e., $\phi$ obeys the Klein--Gordon equation relative to $g$ but $g$ obeys the vacuum Einstein equation). Under the obvious way of understanding this theory, it has the same geometrical degrees of freedom as vacuum general relativity---so it is certainly not anything like background-dependent according to the present approach. But corresponding to each way that the geometry of spacetime could be, there is a vast number of ways of arranging the physical degrees of freedom inherent in the scalar field---so the theory is not anything like background-independent according to the present approach. This is an intuitively attractive verdict: this theory is a sort of hybrid of vacuum general relativity (the paradigm of background-independence) and a theory in which the spacetime metric affects but is not affected by matter (which feature has often been taken to be a hallmark of background-dependence). 

Of course, the theory of the preceding paragraph is a pathological example. At the other end of the spectrum lies general relativity coupled to dust. Given a solution to the Einstein-dust equations, one can reconstruct the dust trajectories and energy density from knowledge of the spacetime metric and the equations of motion 
(see, e.g., \cite[\S 3.14]{Sachs:1977eh}). So if two solutions agree in their geometry, they agree with respect to all physical degrees of freedom---the theory is fully background-independent. 

More typical are cases that lie somewhere in the vast middle ground. The Einstein-Maxwell theory is an interesting example. In effect it splits into a fully background-independent sector and a sector that exhibits the sort of behaviour seen in the hybrid theory above: in generic solutions, one can reconstruct the electromagnetic field from knowledge of the geometry and the equations of motion; but there exist exceptional solutions---those involving null electromagnetic fields---featuring geometries that are compatible with a huge range number of configurations of the electromagnetic field  \cite{Geroch:1966mt}. Here we have a theory that falls short of full background-independence in an illuminating way (and which serves to illustrate the inadequacy of the naive equivalence between a theory's being background-independent and its having as its most natural formulation a generally covariant one). It would be desirable to extend the framework of Section \ref{secBI} to provide comparative measures of background-independence for theories like this one that fall short of near background-independence.

\section*{Acknowledgements}  Thanks to Harvey Brown, John Earman, Michael Friedman, Bob Geroch, David Malament, Brian Pitts, Oliver Pooley, Tom Ryckman, Laura Ruetsche, Rob Rynasiewicz, Ryan Samaroo, Philip Stamp, and Steve Weinstein for helpful comments and conversations. Work on this paper was supported during 2006--2007 by the American Council of Learned Societies and by the Center for Advanced Study in the Behavioral Sciences. 
\bibliographystyle{spmpsci}
\bibliography{gord}
\end{document}